\documentclass[authoryear,preprint,12pt]{elsarticle}

\usepackage{amssymb}
\usepackage[english]{babel}
\usepackage{natbib}
\usepackage{apalike}

\def\smskip{\par\vskip 10pt}
\def\QED{\hfill$\Box$\smskip}

 \usepackage[usenames,dvipsnames]{pstricks}
 \usepackage{epsfig}
\usepackage[displaymath]{lineno}
\usepackage{bm}

\newtheorem{Theorem}{\indent Theorem}[section]
\newtheorem{Lemma}{\indent Lemma}[section]
\newtheorem{Proposition}{\indent Proposition}[section]
\newtheorem{Example}{\indent Example}[section]
\newtheorem{Definition}{\indent Definition}[section]
\newtheorem{Corollary}{\indent Corollary}[section]
\newtheorem{Remark}{\indent Remark}[section]

\newcommand{\proof}{\textbf{Proof: }}

\begin{document}

\bibliographystyle{plainnat}
\begin{frontmatter}

\title{Continuous Representations of Preferences by Means of Two Continuous Functions
}

\author[label1]{Gianni Bosi}
\address[label1]{DEAMS, Universit\`a di Trieste,\\ via
Universit\`a 1, 34123, Trieste, Italy. \\ E-mail:gianni.bosi@deams.units.it}

\author[label2]{Asier Estevan}%
  \address[label2]{{Dpto. Estad\'istica, Inform\'atica y Matem\'aticas,\\ Instituto INAMAT, Universidad P\'ublica de Navarra.  \\ Campus Arrosad\'{\i}a, 31006.  Iru\~na-Pamplona, Navarra, Spain.    }
  } %

\begin{abstract}

Let  $\precsim$ be a reflexive binary relation on a topological space $(X, \tau )$. A  pair $(u,v)$ of continuous real-valued functions on $(X, \tau )$  is said to be a  {\em continuous representation} of $\precsim$  if, for all $x,y \in X$, [$(x \precsim y \Leftrightarrow u(x)  \leq v(y))$]. In this paper we provide a characterization of the existence of a continuous representation of this kind in the general case when neither the functions $u$ and $v$ nor the topological space $(X,\tau )$ are required to satisfy any particular assumptions. Such characterization is based on a suitable continuity assumption of the binary relation $\precsim$, called {\em weak continuity}. In this way, we generalize all the previous results on the continuous representability of interval orders, and also of total preorders, as particular cases.

\end{abstract}

\end{frontmatter}

\section{Introduction}

\indent An interval order $\precsim$ on a set $X$ is a binary relation on $X$ which is \it reflexive \rm and in addition verifies the following condition for all $x,y, z,w \in X$:\linebreak  $[ (x\precsim z) \,\,and \,\,(y \precsim w) \Rightarrow (x\precsim w) \,\,or \,\,(y \precsim z)]$. A reflexive binary relation $\precsim$ on a set $X$ for which there exists a pair $(u,v)$ of  real-valued functions on $X$ such that, for all $x,y \in X$, [$(x \precsim y \Leftrightarrow u(x)  \leq v(y))$], is necessarily an interval order. It is well known that the interest of interval orders in economics is related to the fact that they are not necessarily transitive binary relations, while they are in some sense the simplest ones with this characteristic, since they can be fully represented by means of pairs of real-valued functions in the sense above.

 It is interesting to look for conditions implying the existence of a continuous representation $(u,v)$ of an interval order $\precsim$ on a topological space $(X,\tau )$. Necessary and sufficient conditions for the existence of a continuous representation of an interval order on a connected topological space where presented by \citealp{Cha}. \citealp{Fi2} defined two total preorders $\precsim^{*}$ and $\precsim^{**}$, the so called {\em traces}, which are naturally associated to an interval order $\precsim$. Characterizations of the existence of a continuous representation $(u,v)$ of an interval order on a topological space were presented by  \citealp[Corollary 4.3]{BioCan} and  \citealp[Theorem 1]{BoCanIn} in the case when the traces are required to be continuous, so that the representing functions $u$ and $v$ turn out to be continuous utility functions for the traces. Nevertheless,  a condition of this kind is not necessary for the existence of a continuous representation (see e.g. \citealp[Example 1]{{BoCanIn}}). \\ \indent Necessary conditions for the existence of a continuous representation of an interval order, and respectively a characterization of the existence of a continuous representation when the support set is finite, were presented by \citealp[Proposition 2 and Theorem 6]{IJUFKS2015}.\\ \indent  In this paper we provide a characterization of the existence of a continuous representation of an interval order on a topological space in the general case, i.e. when neither the representing functions nor the topological space are required to satisfy any particular assumptions. Such characterization is based on a suitable continuity assumption of the interval order  $\precsim$, called {\em weak continuity}, which was introduced by \citealp{BoIMF2008}. In this way, we generalize all the previous results on the continuous representability of interval orders, and also of total preorders, as particular cases.  For example, we show that any weakly continuous interval order on a second countable topological space is continuously representable. This result may be viewed as a slight generalization of the famous continuous utility representation theorem of \citealp{De}, according to which a total preorder $\precsim$ on a second countable topological space $(X, \tau )$ admits a continuous utility representation. In the last section we show that our considerations can be easily generalized to the case of {\em biorders} (see \citealp{Doig}).

\section{Notation and preliminary results} \indent  An \it interval order \rm $\precsim$ on an arbitrary nonempty  set $X$ is a binary relation on $X$ which is \it reflexive \rm and in addition verifies the following condition for all $x,y, z,w \in X$: \[ (x\precsim z) \,\,and \,\,(y \precsim w) \Rightarrow (x\precsim w) \,\,or \,\,(y \precsim z).\] \indent  An interval order $\precsim$ on a set $X$ is  {\em total} (i.e., for all $x,y \in X$ we have that either $x \precsim y$ or $y \precsim x$) and not necessarily transitive (see e.g. \citealp{Ol}). The \it irreflexive part \rm of an  interval order $\precsim$ will be denoted by $\prec$ (i.e., for all $x,y \in X$, $x \prec y$ if and only if $(x \precsim y) \mbox{ and not} (y \precsim x)$). \newline  \indent   \citealp{Fi2} proved that if $\precsim$ is an interval order  on a set $X$, then  the following two binary relations $\precsim^{*}$ and $\precsim^{**}$ on $X$ are both total preorders: \[x \precsim^{*} y \Leftrightarrow (z \precsim x \Rightarrow z \precsim y)\,\,\,\mbox{for\,\,\,all}\,\,\,z \in X,\] \[x \precsim^{**} y \Leftrightarrow (y \precsim z \Rightarrow x \precsim z)\,\,\,\mbox{for\,\,\,all}\,\,\,z \in X.\] \indent  Obviously,  if $\precsim$ is a {\em total preorder} $\precsim$ (i.e., $\precsim$ is reflexive, transitive and total), then $\precsim$ is  an interval order. In this case, we have that $\precsim = \precsim^{*}=\precsim^{**}$. The irreflexive parts of $\precsim^{*}$ and $\precsim^{**}$ will be denoted by $\prec^{*}$ and $\prec^{**}$.
Since $\sim^*$ and $\sim^{**}$ are equivalence relations, we may denote the equivalence class of an element $x$ by $\overline{x}$ (since there will not be misunderstanding in the text, we will use the same notation for both equivalence relations).\\
\indent If $R$ is a binary relation on a set $X$, then denote by $L_{R}(x)$ ($U_{R}(x)$) the {\em  lower} ({\em upper}) {\em section} of any element $x \in X$ (i.e., for every $x \in X$, $L_{R}(x)=\{ y \in X: y R x \}$ and  $U_{R}(x)=\{ y \in X: x R y \}$).
\newline \indent We recall that a real-valued function $u$ on a {\em preordered set} $(X, \precsim )$ is said to be {\em increasing} if, for all $x,y \in X$,\[x \precsim y \Rightarrow u(x) \leq u(y).\]\newline \indent    Further, a real-valued function $u$ on $X$ is said to be a {\em utility function} for a total preorder $\precsim$ on a set $X$   if, for all $x,y \in X$,\[x \precsim y \Leftrightarrow u(x) \leq u(y).\]
\indent   A pair $(u,v)$ of real-valued functions on $X$ is said to
{\em represent} an interval order $\precsim$ on $X$ if, for all $x,y
\in X$, \[x \precsim y \Leftrightarrow u(x)  \leq v(y).\] \indent We
say that a pair $(u,v)$ of real-valued functions on $X$  {\em almost
represents} an interval order $\precsim$ on $X$ if, for all $z,w \in
X$, \[(z \precsim w \Rightarrow u(z)  \leq v(w)) \mbox{ and } (z
\prec w \Rightarrow v(z)  \leq u(w)).\]

An interval order $\precsim$ on a topological space $(X,\tau )$ is said to be \it upper (lower) semicontinuous \rm if $L_{\prec}(x)$ ($U_{\prec}(x)$) is an open subset of $X$ for every $x \in X$.  If $\precsim$ is both upper and lower semicontinuous, then it is said to be \it continuous \rm . \\ \indent  Let us now recall the definition of {\em weak continuity} of an interval order on a topological space, which was introduced by \citealp[Definition 2.1]{BoIMF2008} (see also \citealp{BoZuaint2012}).

\begin{Definition}  \label{Defcont} \em An interval order $\precsim$ on a topological space $(X, \tau )$ is {\em weakly continuous} if for every $x,y \in X$ such that $x \prec y$ there exists a pair $(u_{xy}, v_{xy})$ of continuous real-valued functions on $(X, \tau )$ satisfying the following conditions: \begin{enumerate}   \item[(i)] $(u_{xy}, v_{xy})$  almost represents $\precsim$; \\ \item[(ii)] $v_{xy}(x) < u_{xy}(y)$. \end{enumerate} \end{Definition}

The concept of {weak continuity} described in  Definition \ref{Defcont} is reminiscent of the concept of {\em weak continuity} of a preorder on a topological space. From \citealp[Definition 2.3]{hp}, a (not necessarily total) preorder  $\precsim$ on a topological space $(X,\tau )$ is said to be {\em weakly continuous} (see also \citealp{BoHerdSz}) if for every $x, y \in X$ such that $x \prec y$ there exists a continuous increasing real-valued function $u_{xy}$ on $(X, \tau )$ such that\linebreak  $u_{xy}(x) < u_{xy}(y)$.

It is clear that  if there exists a pair $(u,v)$ of continuous real-valued functions   representing an interval order $\precsim$ on $(X,\tau )$, then $\precsim$ is weakly continuous.

\begin{Definition}  \label{Defstrong} \em We say that a total preorder $\precsim$ on a topological space $(X, \tau )$ is {\em almost upper semicontinuous} ({\em almost lower semicontinuous}) if there exists a mapping $L^0_{\prec}: X \rightarrow \tau$ such that, for every $x \in X$, $ x \not \in L^0_{\prec}(x)$, $L^0_{\prec}(x) \supset L_{\prec}(x)$ and $L^0_{\prec}(x)$ is $\prec$-decreasing (respectively. there exists a mapping $U^0_{\prec}: X \rightarrow \tau$ such that, for every $x \in X$, $ x \not \in U^0_{\prec}(x)$, $U^0_{\prec}(x) \supset U_{\prec}(x)$ and $U^0_{\prec}(x)$ is $\prec^{**}$-increasing.).

\end{Definition}

It i clear that an upper (lower) semicontinuous total preorder on a topological space $(X, \tau )$ is almost upper (lower) semicontinuous.

\begin{Lemma} \label{lemmaweakint} Let $\precsim$ be an interval order on a topological space $(X, \tau )$. If $\precsim$ is weakly continuous, then the following conditions hold:

\begin{enumerate}

\item[{\em (i)}] $\precsim$ is continuous;\\

\item[{\em (ii)}] $\precsim^{*}$ is almost lower semicontinuous;\\

\item[{\em (iii)}] $\precsim^{**}$ is almost upper semicontinuous.

 \end{enumerate}
\end{Lemma}

\proof Let $\precsim$ be a weakly continuous interval order on a topological space $(X, \tau )$.

(i). In order to prove that $\precsim$ is continuous, we first  show that $L_{\prec}(x)= \{z \in X : z \prec x\}$ is an open subset of $X$ for every $x \in X$. Consider some fixed $x \in X$, and let $z$ be any element of $X$ such that $z \prec x$. Since $\precsim$ is weakly continuous, there exists an almost representation $(u_{xy},v_{xy})$ of $\precsim$ such that $v_{xy}(z) < u_{xy}(x)$. Therefore, $v_{xy}^{-1}(]- \infty , u_{xy}(x)[)$ is an open subset of $X$ containing $z$ and contained in $L_{\prec}(x)$. Indeed, $x \precsim u$ implies that $u_{xy}(x) \leq v_{xy}(u)$ for every $u \in X$. Analogously it can be shown that $U_{\prec}(x)= \{z \in X : x \prec z\}$ is an open subset of $X$ for every $x \in X$.

(iii). Now let us show that $\precsim^{**}$ is almost upper semicontinuous. Analogously it can be shown that $\prec^{*}$ is almost lower semicontinuous. Define, for every $x \in X$, \begin{eqnarray*} L^0_{\prec^{**}}(x) &=& \bigcup_{(u,v) \in UV_{\precsim}^{ac}} u^{-1}(]- \infty , u(x)[),\end{eqnarray*}
 where $UV_{\precsim}^{ac}$stands for the set of all continuous almost representations of $\precsim$.
It is immediate to check that $x \not \in L^0_{\prec^{**}}(x)$. It is easy to show that $ L^0_{\prec^{**}}(x) \supset  L{\prec^{**}}(x)$, since $z \prec^{**} x$ implies the existence of some point $\xi \in X$ such that $z \precsim \xi \prec x$, so that, from weak continuity of $\precsim$, there exists $(u_{\xi x},v_{\xi x}) \in UV_{\precsim}^{ac}$ with $u_{\xi x}(z) \leq v_{\xi x}(\xi) < u_{\xi x}(x)$. Finally, since it is nearly immediate to realize that $u(z) \leq v(x)$ for every $(u,v) \in UV_{\precsim}^{ac}$, and $z \prec^{**} x$, we have that the above defined set $ L^0_{\prec^{**}}(x)$ is $\prec^{**}$-decreasing. This consideration completes the proof.\QED

The following proposition holds, which reduces weak continuity of a total preorder (see Definition \ref{Defcont}) precisely to
the notion of continuity.

\begin{Proposition} \label{Propweaktotpre} Let $\precsim$ be a total preorder on a topological space $(X, \tau )$. Then the following conditions are equivalent:

\begin{enumerate}
\item[{\em (i)}] $\precsim$ is weakly continuous;\\

\item[{\em (ii)}] $\precsim$ is continuous.

\end{enumerate}

\end{Proposition}

\proof (i) $\Rightarrow$ (ii).  See Lemma \ref{lemmaweakint}.
(ii) $\Rightarrow$ (i). Let $\precsim$ be a continuous total preorder on $(X, \tau )$. From \citealp[Lemma 2.2]{hp}, for every pair $(x,y) \in X \times X$ such that $x \prec y$ there exists a real-valued continuous increasing function $u_{xy}$ such that $u_{xy}(x) < u_{xy}(y)$. Since it is immediate to check that the pair $(u_{xy},u_{xy})$ is a continuous almost representation of $\precsim$, we have that $\precsim$ is weakly continuous. This consideration completes the proof.\QED

\section{Continuous representability}

\indent  We recall the definition of a {scale} in  a topological space and a lemma, the proof of which may be found for example in the proof of the lemma on pages 43-44 in Gillman and Jerison (1960) (see also Theorem 4.1 in  Burgess and  Fitzpatrick (1977)).

\begin{Definition} \label{Defscal} \em If $(X, \tau )$ is a topological space and $\Bbb S$ is a dense subset of $[0,1]$ such that $1 \in {\Bbb S}$, then a family $ \{ G_r \}_{r \in \Bbb S} $ of open subsets of $X$ is  said to be a {\em scale} in $(X, \tau )$ if the following conditions hold:   \item {(i)} $G_1 =X$; \item {(ii)} $\overline{G_{r_1}} \subseteq G_{r_2}$ for every $r_1 , r_2 \in \Bbb S$ such that $r_1 < r_2$. \end{Definition}

\begin{Lemma} \label{Lemmascal} If $ \{ G_r \}_{r \in \Bbb S} $ is a scale in a topological space $(X, \tau )$, then the formula \[ u(x)= inf\{ r \in {\Bbb S}: x \in G_{r} \}\,\,\,\,\,\,\,(x \in X)\] defines a continuous function on $(X,\tau )$ with values in $[0,1]$.  \end{Lemma}

The following characterization of weak continuity  of an interval order on a topological space was proven by \citealp[Proposition 3.2]{BoZuaint2012}.

\begin{Proposition} \label{Propweak} Let  $\precsim $ be an interval order on a topological space $(X,\tau )$. Then the following conditions are equivalent:\\
\item {\em (i)} $\precsim$ is weakly continuous;\\
\item {\em (ii)} For every pair $(x,y) \in X \times X$ such that $x \prec  y$ there exist two scales\linebreak \indent $ \{ G_r^{*(xy)} \}_{r \in \Bbb S} $  and $ \{ G_r^{**(xy)} \}_{r \in \Bbb S} $ in $(X,\tau )$ such that the family\linebreak \indent $\{ (G^{*(xy)}_{r}, G^{**(xy)}_{r}) \}_{r \in \Bbb S}$ satisfies the following conditions: \\ \\ \indent   $(a)$ $z \precsim w$ and $w \in G^{*(xy)}_{r}$ imply $z \in G^{**(xy)}_{r}$ for every $z,w \in X$ and\linebreak \indent \indent \indent  $r \in \Bbb S$;\\ \\ \indent \indent  $(b)$ $z \prec w$ and $w \in G^{**(xy)}_{r}$ imply $z \in G^{*(xy)}_{r}$ for every $z,w \in X$ and\linebreak \indent \indent \indent  $r \in \Bbb S$;  \\  \\ \indent \indent  $(c)$  $x \in G^{*(xy)}_{r}$ and $y \not \in G^{**(xy)}_{r}$ for every $r \in \Bbb S \setminus \{1\}$.\end{Proposition}

We recall that an interval order $\precsim$ on a set $X$ is said to be {\em i.o. separable} if there exists a countable subset $D=\{d_n\}_{n \in {\mathbb N}}$ of $X$ such that for all $x,y \in X$ with $x \prec y$ there exist $d_m , d_n \in D$ such that $x\precsim^*d_m \prec d_n \precsim^{**}y$ (see e.g. \citealp{BoIs2006}). In this case $D$ is said to be an {\em i.o. order dense subset } of the set $X$.

We now present a characterization of the existence of a continuous representation of an interval order on a topological space without imposing any restrictive condition.

\begin{Theorem} \label{TheoGenCar} Let  $\precsim $ be an interval order on a topological space $(X,\tau )$. Then the following conditions are equivalent:

\begin{enumerate}

\item[{\em (i)}] There exists a pair of continuous real-valued functions $(u,v)$ on $(X, \tau )$ representing $\precsim$;\\

\item[{\em (ii)}] There exists a countable family $\{(u_n , v_v )\}_{n \in \mathbb
N \setminus \{0\}}$ of pairs of continuous real-valued functions on
$(X, \tau )$ with values in $[0,1]$ almost representing $\precsim$
such that for every $x , y \in X$ with $x \prec y$ there exists $n
\in \mathbb N \setminus \{0\}$ with $v_n (x) < u_n (y)$;\\

\item[{\em (iii)}] $\precsim$ is i.o. separable and weakly continuous;\\

\item[{\em (iv)}] $\precsim$ is i.o. separable and for every pair $(x,y) \in X \times X$ such that $x \prec  y$ there exist two scales $ \{ G_r^{*(xy)} \}_{r \in \Bbb S} $  and $ \{ G_r^{**(xy)} \}_{r \in \Bbb S} $ in $(X,\tau )$ such that the family $\{ (G^{*(xy)}_{r}, G^{**(xy)}_{r}) \}_{r \in \Bbb S}$ satisfies the following conditions:\\ \begin{enumerate} \item[{\em (a)}] $z \precsim w$ and $w \in G^{*(xy)}_{r}$ imply $z \in G^{**(xy)}_{r}$ for every $z,w \in X$ and  $r \in \Bbb S$;\\ \item[{\em (b)}]    $z \prec w$ and $w \in G^{**(xy)}_{r}$ imply $z \in G^{*(xy)}_{r}$ for every $z,w \in X$ and $r \in \Bbb S$;\\\item[{\em (c)}]  $x \in G^{*(xy)}_{r}$ and $y \not \in G^{**(xy)}_{r}$ for every $r \in \Bbb S \setminus \{1\}$;\\\end{enumerate}

\end{enumerate}

\end{Theorem}

\proof (i) $\Rightarrow$ (ii). Immediate.\\
\medskip
(ii)$\Rightarrow$ (i). Since the extended real line is homeomorphic to $[0,1]$, we may assume that functions $u_n$ and $v_n$ take values on $[0,1]$, for every $n\in \mathbb{N}$. Hence, it is straithgforward  to see that the pair of functions $(u,v)$ defined by $u=\sum\limits_1^\infty \frac{u_n}{2^n}$ and $v=\sum\limits_1^\infty \frac{v_n}{2^n}$ is a continuous representation of the interval order.\\
\medskip

(ii) $\Rightarrow$ (iii). \citealp[Proposition 2.1]{BoZuaint2012} actually proved that it is also the case that the above condition (ii) implies the existence of a continuous representation $(u,v)$ for the interval order $\precsim$. Therefore, condition (iii) immediately descends.\\
\medskip

(iii) $\Rightarrow$ (ii). Since the interval order is i.o.-separable, there exists a countable set $D\subseteq X$ such that for any $x\prec y $ there exist $d_n,d_m\in D$ such that $x\precsim^* d_n\prec d_m\precsim^{**} y.$ Since the interval order is weakly continuous, there exists a (possible uncountable) family $\{(u_i,v_i)\}_{i\in I}$ of continuous almost representations such that for any pair $x\prec y$ there is an index $i\in I$ such that $v_i(x)<u_i(y).$ Let' see that there also exists a countable family $\{(u_j,v_j)\}_{j\in J}$ that satisfies those conditions.

We argue on a countable set $\overline{D}$ that extends $D$. For any $x\prec y $ there exist $d_n,d_m\in D$ such that $x\precsim^* d_n\prec d_m\precsim^{**} y,$ so --since $d_n\prec d_m$-- there exists a pair of functions $(u_{nm}, v_{nm})$ such that $v_{nm}(d_n)<u_{nm}(d_m)$. If it holds that   $x\prec^* d_n\prec d_m\prec^{**} y,$ then $v_{nm}(x)\leq v_{nm}(d_n)<u_{nm}(d_m)\leq u_{nm}(y)$ is satisfied, so the inequality also holds true for the pair $x\prec y$.
However, if $x\sim^*d_n$ or $y\sim^{**} d_m$, the inequality $v_{nm}(x)<u_{nm}(y)$ may fail to be true.
In case $x\sim^* d_n$, we distinguish two cases:
\begin{enumerate}
\item[$(a)$] If there exists $d_n'\in \overline{d_n}$ such that $v_{nm}(d_n')=\max \{v_{nm}(d)\colon d\in \overline{d_n}\}$, then we add $d_n'$ to the initial set $\overline{D}$ that extends $D$. Notice that, since $d_n'\sim^*d_n \prec d_m$, then it holds $d_n'\prec d_m$ too. Dually, if there exists $d_m'\in \overline{d_m}$ such that $u_{nm}(d_m')=\min \{u_{nm}(d)\colon d\in \overline{d_m}\}$, then we add $d_m'$ too.

\item[$(b)$] If there is no maximum (case (a)), then there is a supremum  of $\{v_{nm}(d)\colon d\in \overline{d_n}\}$. Hence, there exists a sequence $(d_n^k)_{k\in \mathbb{N}}\subseteq \overline{d_n}$ such that $(v_{nm}(d_n^k))_{k\in \mathbb{N}}$ is strictly increasing (i.e. $v_{nm}(d_n^1)<v_{nm}(d_n^2)<\cdots <v_{nm}(d_n^k)<\cdots$) and it converges to the supremum. In particular, notice that there is an index $k_0\in \mathbb{N}$ such that $v_{nm}(x)<v_{nm}(d_n^k)$ for every $k>k_0$. In that case, we add all these elements of the sequence $(d_n^k)_{k\in \mathbb{N}}\subseteq \overline{d_n}$ (so, a countable number of elements) to $\overline{D}$. Again, notice that, since $d_n^k\sim^*d_n \prec d_m$ (for any index $k$), then it holds $d_n^k\prec d_m$ too. We reason dually with the infimum in case there is not a minimum $u_{nm}(d_m')=\min \{u_{nm}(d)\colon d\in \overline{d_m}\}$ (case (a)).
\end{enumerate}

Therefore, now we have a countable subset $\overline{D}\subseteq X$ such that for any pair $x\prec y$, there exist $d_n,d_m\in \overline{D}$ such that $x\precsim^* d_n \prec d_m \precsim^{**} y$ and satisfying that there exists a pair of functions $(u_{nm}, v_{nm})$ such that $v_{nm}(d_n)<u_{nm}(d_m)$ as well as $v_{nm}(x)\leq v_{nm}(d_n)<u_{nm}(d_m)\leq u_{nm}(y)$.
Thus we just need those pair of functions $(u_i,v_i)$ that distinguish the elements of $\overline{D}$, i.e. a countable number of functions $\{(u_{nm}, v_{nm})\}_{n,m\in \mathbb{N}}$.

\medskip
(iii) $\Rightarrow$ (iv). See \citealp[Proposition 3.2]{BoZuaint2012}

\medskip

(iv) $\Rightarrow$ (ii). Based on Lemma \ref{Lemmascal} and Proposition \ref{Propweak}, it is now nearly immediate to establish this implication.
\QED

\begin{Corollary} \label{CorDeb} Let $\precsim$ be a weakly  interval order on a second countable topological space $(X, \tau )$. Then there exists a pair of continuous real-valued functions $(u,v)$ on $(X, \tau )$ representing $\precsim$.\end{Corollary}

\proof From \citealp[Proposition 2.3]{Br3} and Lemma \ref{lemmaweakint}, (i), there exists a representation $(u',v')$ of the interval order $\precsim$, so that Theorem \ref{TheoGenCar} applies.\QED

It is now easy to check that Corollary \ref{CorDeb} generalizes Debreu theorem.

\begin{Corollary}[Debreu theorem] \label{Debreu} Let $\precsim$ be a continuous total preorder on a second countable topological space $(X, \tau )$. Then there exists a continuous utility function $u$  on $(X, \tau )$ representing $\precsim$.\end{Corollary}

\proof From Proposition \ref{Propweaktotpre}, we have that $\precsim$ is weakly continuous, so that it is continuously representable by  a pair $(u,v)$ of real-valued functions by Corollary \ref{CorDeb}. However, from the proof of Proposition \ref{Propweaktotpre}, it is clear that we can consider $u_n = v_n$ for every $n \in {\Bbb N}^+$ in condition (ii) of Theorem \ref{TheoGenCar}, so that actually we can take $u=v$, a continuous utility representation for $\precsim$. \QED

Finally let's see that 
i.o.-separability is not a necessary condition for the existence of a continuous almost representation (of course, without it a continuous representation cannot exist).
\begin{Example}\label{notrep}\rm
Let $\precsim$ be the lexicographic total order defined on $X=(0,1)\times (0,1)$:
$$
 (a,b) \prec (x, y)\iff  \left\{
  \begin{array}{ll}
   a< x  &\mbox{ ; }  \forall b,y. \\[4pt]
    a= x  &\mbox{ ; } b<y.\\[4pt]
  \end{array}\right.
$$
So, $(a,b)\sim (x,y) $ if and only if $ a=x$ and  $b=y$. We endow $X$ with  the order topology induced by the lexicographic order.

It is well-known (see Bridges and Mehta \cite{BRME}) that the lexicographic total order on the plane fails to be perfectly-separable and, hence, it is not representable by means of a utility function $u$ such that $(a,b)\prec (x,y)\iff u((a,b)<u((x,y)).$

Let $\pi$ be the projection on the first coordinate, i.e.  $\pi((x,y))=x$ for any $(x,y)\in X$. Now, for each $r\in (0,1)$ we define the following  function on $X$:
\[
u_r((x,y))=  \left\{
  \begin{array}{ll}
   0  &\mbox{ ; } x<r.\\[4pt]
   y  &\mbox{ ; } x=r.\\[4pt]
   1 &\mbox{ ; } r<x.\\[4pt]
  \end{array}\right.
\]

Thus, functions $u_r$ (for every $r\in (0,1)$) and $\pi $ are almost representations of the total order, since they satisfy that
\begin{enumerate}
\item $(a,b)\precsim (x,y) \Rightarrow v((a,b))\leq v((x,y)),$
\item $(a,b)\prec (x,y) \Rightarrow v((a,b))\leq v((x,y)).$
\end{enumerate}
Moreover, it can be proved that $u_r$ is continuous (as well as   projection $\pi$) with respect to the order topology induced by the lexicographic order. Hence, we conclude that the order is weakly continuous, since for any two points $(a,b)$ and $(x,y)$ with $(a,b)\prec (x,y)$ it holds that there exists a continuous almost representation $v$ satisfying that $v((a,b))<v((x,y))$ (in fact, $v=\pi$ if $a<x$ and $v=u_a$ if $a=x$).

 \end{Example}

\section{Continuous representability of biorders}

As a consequence of the study before for interval orders, we are also able to include some results concerning to the continuous representability of biorders. First, we present some key definitions (see \cite{Doig}).

\begin{Definition} \rm
A binary relation $\prec$ from $A$ to $X$ is a {\em biorder} if it is Ferrers, that is, if for
every $a, b \in A$ and $x, y \in X$ the 
$\{a\prec x) $ and $(b\prec y) \}$ implies that $\{(a\prec y) $ or $(b\prec x)\}.$

As usual, we denote  by $x\precsim a $ whenever $\neg (a\prec x)$, $a\in A,$ $x\in X$.
\end{Definition}

\begin{Definition} \rm
A biorder $\prec$ from $A$ to $X$  is
said to be {\em representable} (as well as {\em realizable with respect to $<$}, see \cite{Doig}) if there exists a pair of  real-valued functions $v \colon A \to \mathbb{R}$, $u \colon X \to \mathbb{R}$  such that $a\prec x \iff v(a)<u(x)$, $a\in A$, $x\in X$.

It is said to be {\em representable with respect to $\leq$} (see \cite{Doig}) if there exists a pair of  real-valued functions $v \colon A \to \mathbb{R}$, $u \colon X \to \mathbb{R}$  such that $a\prec x \iff v(a)\leq u(x)$, $a\in A$, $x\in X$.
\end{Definition}

Furthermore, we may extend the notion of {\em almost representability} used for interval orders to biorders in a natural manner:

\begin{Definition}\rm
We say that a pair $(v,u)$ of real-valued functions on $A$ and $X$, respectively, {\em almost
represents} a biorder $\precsim$ from $A$ to $X$ if, for all $a\in A$ and $x \in
X$, $(x \precsim a \Rightarrow u(x)  \leq v(a)) \mbox{ and } (a
\prec x \Rightarrow v(a)  \leq u(x)).$
\end{Definition}

\begin{Remark}
Notice that the concept of almost representability for biorders is weaker than the representability with respect to $\leq$, as well as weaker than the representability with respect to $<$.
\end{Remark}

The following characterization is known in the literature (see \cite{Doig} as well as \cite{Naka}).

\begin{Theorem}
A biorder $\prec$ from $A$ to $X$ is representable (with respect to $<$) if and only if it is {\em jointly dense}, that is, there exists a countable set $D=A'\cup X'$ (with $A'\subseteq A$ and $X' \subseteq X$) such that for any pair $a\prec x$ there exist $b\in A'$ and $y\in X'$ such that $a\precsim^* b\prec y\precsim^{**} x$.\footnote{Here, as for interval orders, $\precsim^*=\prec \circ \precsim $ and $\precsim^{**}=\precsim\circ \prec$.}
\end{Theorem}

Hence, now we may generalize the concept of weakly continuous representation for biorders.

\begin{Definition}  \label{Defcontbi}\rm  A  biorder $\prec$ from a topological space $(A,\tau_A)$ to another one $(X, \tau_X )$ is {\em weakly continuous} if for every $a\in A$ and $x \in X$ such that $a \prec x$ there exists a pair $(v_{ax}, u_{ax})$ of continuous real-valued functions on $(A,\tau_A)$ and $(X, \tau )$ (respectively) satisfying the following conditions:
 \begin{enumerate}   \item[(i)] $(v_{ax}, u_{ax})$  almost represents $\prec$; \\ \item[(ii)] $v_{ax}(a) < u_{ax}(x)$. \end{enumerate} \end{Definition}

Finally, we present as a corollary the following result, which proof is reduced to a sketch since it is similar to that developed for interval orders.

\begin{Theorem} \label{TheoGenCarB} Let  $\prec$ be a biorder from a topological space $(A,\tau_A )$ to $(X,\tau_X)$. Then the following conditions are equivalent:

\begin{enumerate}

\item[{\em (i)}] There exists a pair of continuous real-valued functions $(v,u)$ on $(A, \tau_A) $ and $(X, \tau_X )$, respectively, representing $\prec$ with respect to $<$;\\

\item[{\em (ii)}] There exists a countable family $\{(v_n , u_n )\}_{n \in \mathbb
N \setminus \{0\}}$ of pairs of continuous real-valued functions on
$(A, \tau_A)$ and $(X, \tau_X )$ (respectively) with values in $[0,1]$ almost representing $\prec$
such that for every $a\in A$ and $x  \in X$ with $a \prec x$ there exists $n
\in \mathbb N \setminus \{0\}$ with $v_n (a) < u_n (x)$;\\

\item[{\em (iii)}] $\prec$ is jointly-dense and weakly continuous.

\end{enumerate}

\end{Theorem}

\proof (i) $\Rightarrow$ (ii). Immediate.\\
\medskip
(ii)$\Rightarrow$ (i). As we did for interval orders, here again the pair of functions $(v,u)$ defined by $u=\sum\limits_1^\infty \frac{u_n}{2^n}$ and $v=\sum\limits_1^\infty \frac{v_n}{2^n}$ is a continuous representation of the biorder.\\
\medskip

(ii) $\Rightarrow$ (iii). The weakly continuity is trivial by (ii). Siince (ii) implies (i) and this last one implies to be jointly dense, implication (ii) $\Rightarrow$ (iii) is clear.
\medskip

(iii) $\Rightarrow$ (ii). Since the biorder is jointly-dense, there exists a countable set $D=A'\cup X'$ (with $A'\subseteq A$ and $X' \subseteq X$) such that for any pair $a\prec x$ there exists $b\in A'$ and $y\in X'$ such that $a\precsim^* b\prec y\precsim^{**} x$. Since the biorder is weakly continuous, there exists a (possible uncountable) family $\{(v_i,u_i)\}_{i\in I}$ of continuous almost representations such that for any pair $a\prec x$ there is an index $i\in I$ such that $v_i(a)<u_i(x).$
The proof consist on finding a countable family $\{(v_j,u_j)\}_{j\in J}$ that satisfies those conditions. This is done as for the case of interval orders, extending the countable set $D=A'\cup X'$ to  a bigger countable one $\overline{D'}=\overline{A'}\cup \overline{X'}$, again distinguishing two cases depending on the existence of maximum and minimum values.

After the construction of that countable set $\overline{D'}=\overline{A'}\cup \overline{X'}$, we only need those pair of functions $(v_i,u_i)$ that distinguish the elements of $\overline{A'}$ from $\overline{X'}$, i.e. a countable number of functions $\{(u_{nm}, v_{nm})\}_{n,m\in \mathbb{N}}$.
\QED

\end{document}